\documentclass[twocolumn,showpacs,preprintnumbers,amsmath,amssymb]{revtex4}%
\usepackage{graphicx}%
\usepackage{amsfonts}%

\begin{document}

\title{The Extended Plane Wave Expansion Method in Three Dimensional Anisotropic
Photonic Crystal}

\author{Young-Chung Hsue}
 \email{ychsu@northwestern.edu}
  \affiliation{Department of Physics and Astronomy, Northwestern Unversity, Evanston, Illinois 60201}
\author{Ben-Yuan Gu}
 \email{guby@aphy.iphy.ac.cn}
  \affiliation{ Institute of Physics, Academia Sinica, P.O. Box 603, Beijing 100080, China}

\begin{abstract}
In this paper, we extend the conventional plane wave expansion method in 3D
anisotropic photonic crystal to be able to calculate the complex $\mathbf{k}$
even if permittivity and permeability are complex numbers or the functions of
$\omega$. There are some tricks in the derivation process, so we show the
process in detail. Besides, we also provide an example for testing and
explaining, and we also compare the results with the band structure derived
from conventional plane wave expansion method, then we finally find that there is a
good consistency between them.

\end{abstract}
\maketitle
Recently, the researches of the properties of the photonic crystals (PCs) have aroused great interests, since the concept of the PCs has been proposed by Yablonovitch and John\cite{1,2,3}. Briefly speaking, PCs are periodically structured electromagnetic media, generally processing photonic band gap (PBG). Most of the studies stress the PBG structures with the use of conventional plane-wave expanded (PWE) method\cite{7,8}. However, there are still many articles explore the influence of interface, such as the studies of transmission, reflection, and the penetration depth etc.\cite{9,10,11,12} Furthermore, the penetration depth relates to the imaginary part of wave vector.
As for the complex ${\bf k}$ calculation in 2D isotropic photonic crystals, we had sufficiently discussed
about it in the last paper[13]. Now, this paper is to continue with the last
one. Furthermore, the emphasis of this paper is put on the general formula, 3D
anisotropic case, of extended plane wave expansion (EPWE) method.

Though the main part of the idea resembles in 2D isotropic case[13], the
formula and derivative process are much more complicated than that in 2D
isotropic case, because the basis of wave functions can not be treated as
scalar functions, TE and TM modes in 2D isotropic case. However, the problem
of the difficult part has been overcome and we will explain it in the
following description.
Besides, the eigenfunctions set derived from this EPWE method is completely
the same as that derived from the conventional PWE method. So we have no qualms
about the inaccuracy of the propagation modes between these two methods. 

The system we discussed is periodically structured without charge $\rho$ and
current $\mathbf{J}$.
Therefore, according to Maxwell Equation, the magnetic field $\mathbf{H}%
(\mathbf{r})$ should obey
\begin{align}
-\left(  \mathbf{k+G}\right)  \times\hat{\epsilon}_{\mathbf{G-G}^{\prime}%
}^{-1}\left(  \mathbf{k+G}^{\prime}\right)  \times\mathbf{H}_{\mathbf{G}%
^{\prime}}=\omega^{2}\hat{\mu}_{\mathbf{G-G}^{\prime}}\mathbf{H}%
_{\mathbf{G}^{\prime}},
\end{align}
where
\begin{align*}
\mathbf{H}(\mathbf{r})  &  =\sum\limits_{\mathbf{G}}\mathbf{H}_{\mathbf{G}%
}e^{i(\mathbf{k+G})\cdot\mathbf{r}},\\
\hat{\epsilon}(\mathbf{r})  &  =\sum\limits_{\mathbf{G}}\hat{\epsilon
}_{\mathbf{G}}e^{i\mathbf{G}\cdot\mathbf{r}},\\
\hat{\mu}(\mathbf{r})  &  =\sum\limits_{\mathbf{G}}\hat{\mu}_{\mathbf{G}%
}e^{i\mathbf{G}\cdot\mathbf{r}},
\end{align*}
$\mathbf{G}$ and $\mathbf{G}^{\prime}$ are the reciprocal lattice vectors,
$\omega$ and $\mathbf{k}$ are the frequency and wave vector, $\hat{\epsilon
}(\mathbf{r})$ and $\hat{\mu}(\mathbf{r})$ are the tensors of permittivity and
permeability of which $\hat{\epsilon}_{\mathbf{G}}$ and $\hat{\mu}%
_{\mathbf{G}}$ are the Fourier expansion components, respectively.

Now, let us expand Eq.(1) directly through $\mathbf{\hat{x}}$%
,$\mathbf{\hat{y}}$ and $\mathbf{\hat{z}}$ directions%
\begin{widetext}
\begin{subequations}
\begin{align}
\mathbf{\hat{x}}  &  \mathbf{:} \notag \\ \notag
&  \left[  \left(  k+G\right)  _{y}\epsilon_{zy}^{-1}\left(  k+G^{\prime
}\right)  _{z}+\left(  k+G\right)  _{z}\epsilon_{yz}^{-1}\left(  k+G^{\prime
}\right)  _{y}-\left(  k+G\right)  _{y}\epsilon_{zz}^{-1}\left(  k+G^{\prime
}\right)  _{y}-\left(  k+G\right)  _{z}\epsilon_{yy}^{-1}\left(  k+G^{\prime
}\right)  _{z}\right]  H_{x}\\ \notag
&  +\left[  \left(  k+G\right)  _{y}\epsilon_{zz}^{-1}\left(  k+G^{\prime
}\right)  _{x}+\left(  k+G\right)  _{z}\epsilon_{yx}^{-1}\left(  k+G^{\prime
}\right)  _{z}-\left(  k+G\right)  _{y}\epsilon_{zx}^{-1}\left(  k+G^{\prime
}\right)  _{z}-\left(  k+G\right)  _{z}\epsilon_{yz}^{-1}\left(  k+G^{\prime
}\right)  _{x}\right]  H_{y}\\ \notag
&  +\left[  \left(  k+G\right)  _{y}\epsilon_{zx}^{-1}\left(  k+G^{\prime
}\right)  _{y}+\left(  k+G\right)  _{z}\epsilon_{yy}^{-1}\left(  k+G^{\prime
}\right)  _{x}-\left(  k+G\right)  _{y}\epsilon_{zy}^{-1}\left(  k+G^{\prime
}\right)  _{x}-\left(  k+G\right)  _{z}\epsilon_{yx}^{-1}\left(  k+G^{\prime
}\right)  _{y}\right]  H_{z}\\ 
&  =-\omega^{2}\left(  \mu_{xx}H_{x}+\mu_{xy}H_{y}+\mu_{xz}H_{z}\right)  ,
\end{align}%
\begin{align}
\mathbf{\hat{y}}  &  \mathbf{:} \notag \\ \notag
&  \left[  \left(  k+G\right)  _{z}\epsilon_{xy}^{-1}\left(  k+G^{\prime
}\right)  _{z}+\left(  k+G\right)  _{x}\epsilon_{zz}^{-1}\left(  k+G^{\prime
}\right)  _{y}-\left(  k+G\right)  _{z}\epsilon_{xz}^{-1}\left(  k+G^{\prime
}\right)  _{y}-\left(  k+G\right)  _{x}\epsilon_{zy}^{-1}\left(  k+G^{\prime
}\right)  _{z}\right]  H_{x}\\ \notag
&  +\left[  \left(  k+G\right)  _{z}\epsilon_{xz}^{-1}\left(  k+G^{\prime
}\right)  _{x}+\left(  k+G\right)  _{x}\epsilon_{zx}^{-1}\left(  k+G^{\prime
}\right)  _{z}-\left(  k+G\right)  _{z}\epsilon_{xx}^{-1}\left(  k+G^{\prime
}\right)  _{z}-\left(  k+G\right)  _{x}\epsilon_{zz}^{-1}\left(  k+G^{\prime
}\right)  _{x}\right]  H_{y}\\ \notag
&  +\left[  \left(  k+G\right)  _{z}\epsilon_{xx}^{-1}\left(  k+G^{\prime
}\right)  _{y}+\left(  k+G\right)  _{x}\epsilon_{zy}^{-1}\left(  k+G^{\prime
}\right)  _{x}-\left(  k+G\right)  _{z}\epsilon_{xy}^{-1}\left(  k+G^{\prime
}\right)  _{x}-\left(  k+G\right)  _{x}\epsilon_{zx}^{-1}\left(  k+G^{\prime
}\right)  _{y}\right]  H_{z}\\
&  =-\omega^{2}\left(  \mu_{yx}H_{x}+\mu_{yy}H_{y}+\mu_{yz}H_{z}\right)  ,
\end{align}%
\begin{align}
\mathbf{\hat{z}}  &  \mathbf{:} \notag \\ \notag
&  \left[  \left(  k+G\right)  _{x}\epsilon_{yy}^{-1}\left(  k+G^{\prime
}\right)  _{z}+\left(  k+G\right)  _{y}\epsilon_{xz}^{-1}\left(  k+G^{\prime
}\right)  _{y}-\left(  k+G\right)  _{x}\epsilon_{yz}^{-1}\left(  k+G^{\prime
}\right)  _{y}-\left(  k+G\right)  _{y}\epsilon_{xy}^{-1}\left(  k+G^{\prime
}\right)  _{z}\right]  H_{x}\\ \notag
&  +\left[  \left(  k+G\right)  _{x}\epsilon_{yz}^{-1}\left(  k+G^{\prime
}\right)  _{x}+\left(  k+G\right)  _{y}\epsilon_{xx}^{-1}\left(  k+G^{\prime
}\right)  _{z}-\left(  k+G\right)  _{x}\epsilon_{yx}^{-1}\left(  k+G^{\prime
}\right)  _{z}-\left(  k+G\right)  _{y}\epsilon_{xz}^{-1}\left(  k+G^{\prime
}\right)  _{x}\right]  H_{y}\\ \notag
&  +\left[  \left(  k+G\right)  _{x}\epsilon_{yx}^{-1}\left(  k+G^{\prime
}\right)  _{y}+\left(  k+G\right)  _{y}\epsilon_{xy}^{-1}\left(  k+G^{\prime
}\right)  _{x}-\left(  k+G\right)  _{y}\epsilon_{xx}^{-1}\left(  k+G^{\prime
}\right)  _{y}-\left(  k+G\right)  _{x}\epsilon_{yy}^{-1}\left(  k+G^{\prime
}\right)  _{x}\right]  H_{z}\\
&  =-\omega^{2}\left(  \mu_{zx}H_{x}+\mu_{zy}H_{y}+\mu_{zz}H_{z}\right)  ,
\end{align}
\end{subequations}
\end{widetext}
where $\epsilon_{ij}$ and $\mu_{ij}$ are the abbreviations of $\hat{\epsilon}_{\mathbf{G-G}^{\prime}%
,i,j}$ and  $\hat{\mu}_{\mathbf{G-G}^{\prime},i,j}$, and $H_i$ is the abbreviation of $H_{\mathbf{G}^{\prime},i}$.
When $\mathbf{k}$ is provided, Eq.(2) becomes an eigenvalue problem in
which the eigenvalue is $\omega$ and is the conventional PWE 
method. Now, there comes up an interesting question that is
whether $\mathbf{k}$ must be a vector of which the components are real
numbers. The answer is ''No'', and we just need to do some modification on
Eq.(2) to get the complex $\mathbf{k}$, because Eq.(2) is a 4 variables
($\mathbf{k}$ and $\omega$) equation.

In the beginning, two important things need discussing. First, the inner
product of $(\mathbf{k+G})$ and Eq.(2) results in
\[
\sum\limits_{\substack{G^{\prime},i,j\\i,j=x,y,z}}(k_{i}+G_{i})\mu
_{\mathbf{G-G}^{\prime},i,j}H_{\mathbf{G}^{\prime},j}=0
\]
which are the restriction functions of which the amount is $N$, meanwhile, $N$
is the amount of $\{G\}$ set.
Therefore, the certain amount of the independent eigenfunctions in Eq.(2) is
$2N$ not $3N$. That's why we will get the fake eigenvalues which are
$\omega^{2}=0$ if Eq.(2) is calculated as an eigenvalue equation directly.

To avoid this situation occurring in our method, the eigenvector we selected
in our method is $\left(
\begin{array}
[c]{c}%
\mathbf{H}_{\mathbf{G}\perp}\\
\mathbf{\tilde{H}}_{\mathbf{G}\perp}%
\end{array}
\right)  $ not $\left(
\begin{array}
[c]{c}%
\mathbf{H}_{\mathbf{G}}\\
\mathbf{\tilde{H}}_{\mathbf{G}}%
\end{array}
\right)  $, where $\mathbf{H}_{\mathbf{G}\perp}$ and $\mathbf{H}_{\mathbf{G}}$
are $\left(
\begin{array}
[c]{c}%
H_{\mathbf{G},y}\\
H_{\mathbf{G},z}%
\end{array}
\right)  $ and $\left(
\begin{array}
[c]{c}%
H_{\mathbf{G},x}\\
\mathbf{H}_{\mathbf{G}\perp}%
\end{array}
\right)  $, $\mathbf{\tilde{H}}_{\mathbf{G}\perp}$ and $\mathbf{\tilde{H}%
}_{\mathbf{G}}$ are $k_{x}\mathbf{H}_{\mathbf{G}\perp}$ and $k_{x}%
\mathbf{H}_{\mathbf{G}}$, respectively.

Second, there are no $k_{x}^{2}H_{\mathbf{G}^{\prime},i}$ , $i=x,y,z$, and
$k_{x}H_{\mathbf{G}^{\prime},x}$ in Eq.(2a), which is the $\mathbf{\hat{x}}$
component of Eq.(1), because the inner products of Eq.(1) and $\mathbf{\hat
{x}}$ will cause the existence of just one $k_{x}$ or even no, and
$(\mathbf{k}+\mathbf{G}\prime)\times\mathbf{H}_{\mathbf{G\prime}}$ part will
restrict the existence of $k_{x}H_{\mathbf{G,}x}$.

\bigskip Therefore, the treatment of $\mathbf{\hat{x}}$ component will be
different from $\mathbf{\hat{y}}$ and $\mathbf{\hat{z}}$ components. The
following is the detail derivation process:

\bigskip

First of all, the $\mathbf{\hat{y}}$ and $\mathbf{\hat{z}}$ components of
Eq.(2) can be written as a matrix formula%
\[
\left[  \mathbf{\hat{B}}_{1}\vdots\mathbf{\hat{B}}_{2}\vdots\mathbf{\hat{C}}%
_{1}\vdots\mathbf{\hat{C}}_{2}\right]  \left(
\begin{array}
[c]{c}%
\mathbf{H}_{\mathbf{G}}\\
\mathbf{\tilde{H}}_{\mathbf{G}}%
\end{array}
\right)  =\mathbf{\hat{A}}\left(  k_{x}\mathbf{\tilde{H}}_{\mathbf{G\perp}%
}\right)  ,
\]

and its expansion type is%
\begin{align}
\mathbf{\hat{B}}_{1}H_{\mathbf{G},x}+\mathbf{\hat{C}}_{1}\tilde{H}%
_{\mathbf{G},x}+\left(  \mathbf{\hat{B}}_{2}\vdots\mathbf{\hat{C}}_{2}\right)
\left(
\begin{array}
[c]{c}%
\mathbf{H}_{\mathbf{G}\perp}\\
\mathbf{\tilde{H}}_{\mathbf{G}\perp}%
\end{array}
\right)  =k_{x}\mathbf{\hat{A}}\left(  \mathbf{\tilde{H}}_{\mathbf{G\perp}%
}\right)  ,
\end{align}
where $\mathbf{\hat{A}}$, $\mathbf{\hat{B}}_{1}$, $\mathbf{\hat{B}}_{2}$,
$\mathbf{\hat{C}}_{1}$, $\mathbf{\hat{C}}_{2}$ are $2N\times2N$, $2N\times N$,
$2N\times2N$, $2N\times N$ and  $2N\times2N$ matrices and their elements will
be illustrated in Appendix.

As regards the $\mathbf{\hat{x}}$ component of Eq.(2), we can write in another
form which is different from $\mathbf{\hat{y}}$ and $\mathbf{\hat{z}}$
components of Eq.(2). Thus the matrix form of Eq.(2a) is
\begin{align}
\left(  \mathbf{\hat{E}}_{1}\vdots\mathbf{\hat{E}}_{2}\right)  \left(
\mathbf{H}_{\mathbf{G}}\right)  =\mathbf{\hat{D}}\left(  \mathbf{\tilde{H}%
}_{\mathbf{G\perp}}\right)  ,
\end{align}
where $\mathbf{\hat{D}}$, $\mathbf{\hat{E}}_{1}$, $\mathbf{\hat{E}}_{2}$ are
$N\times2N$, $N\times N$, $N\times2N$ and their elements are also in Appendix.

From Eq.(4) we obtain
\begin{subequations}
\begin{align}
H_{\mathbf{G},x}  & =-\mathbf{\hat{E}}_{1}^{-1}\mathbf{\hat{E}}_{2}%
\mathbf{H}_{\mathbf{G\perp}}+\mathbf{\hat{E}}_{1}^{-1}\mathbf{\hat{D}\tilde
{H}}_{\mathbf{G\perp}},\\
\tilde{H}_{\mathbf{G},x}  & =-\mathbf{\hat{E}}_{1}^{-1}\mathbf{\hat{E}}%
_{2}\mathbf{\tilde{H}}_{\mathbf{G\perp}}+k_{x}\mathbf{\hat{E}}_{1}%
^{-1}\mathbf{\hat{D}\tilde{H}}_{\mathbf{G\perp}},
\end{align}
\end{subequations}
where Eq.(5b) is the production of Eq.(5a) multiplied by $k_{x}$. A
combination of Eqs.(3) and (5) yields
\begin{widetext}
\begin{align*}
k_{x}\mathbf{\tilde{H}}_{\mathbf{G\perp}} &  =\left[  \mathbf{\hat{A}-\hat{C}%
}_{1}\mathbf{\hat{E}}_{1}^{-1}\mathbf{\hat{D}}\right]  ^{-1}\left[  \left(
\mathbf{\hat{B}}_{1}\mathbf{\hat{E}}_{1}^{-1}\right)  \left(  -\mathbf{\hat
{E}}_{2}\vdots\mathbf{\hat{D}}\right)  +\left(  \mathbf{\hat{C}}%
_{1}\mathbf{\hat{E}}_{1}^{-1}\right)  \left(  \O\vdots-\mathbf{\hat{E}%
}_{2}\right)  +\left(  \mathbf{\hat{B}}_{2}\vdots\mathbf{\hat{C}}_{2}\right)
\right]  \left(
\begin{array}
[c]{c}%
\mathbf{H}_{\mathbf{G}\perp}\\
\mathbf{\tilde{H}}_{\mathbf{G}\perp}%
\end{array}
\right)  \\
& \equiv \mathbf{\hat{F}}\left(
\begin{array}
[c]{c}%
\mathbf{H}_{\mathbf{G}\perp}\\
\mathbf{\tilde{H}}_{\mathbf{G}\perp}%
\end{array}
\right)  ,
\end{align*}
\end{widetext}
where $\mathbf{\hat{F}}$ is a $2N \times 4N$ matrix.
Considering the equation above with $k_{x}\mathbf{H}_{\mathbf{G\perp}%
}=\mathbf{\tilde{H}}_{\mathbf{G\perp}}$, we finally have an equation
\begin{align}
\left(
\begin{array}
[c]{c}%
\; \O \; \vdots \; \mathbf{I} \; \\
\cdots\cdots\\
\mathbf{F}%
\end{array}
\right)  \left(
\begin{array}
[c]{c}%
\mathbf{H}_{\mathbf{G}\perp}\\
\mathbf{\tilde{H}}_{\mathbf{G}\perp}%
\end{array}
\right)  =k_{x}\left(
\begin{array}
[c]{c}%
\mathbf{H}_{\mathbf{G}\perp}\\
\mathbf{\tilde{H}}_{\mathbf{G}\perp}%
\end{array}
\right)  ,
\end{align}
which is an $k_{x}$ eigenvalue equation, and the order of eigenfunction
$\left(
\begin{array}
[c]{c}%
\mathbf{H}_{\mathbf{G}\perp}\\
\mathbf{\tilde{H}}_{\mathbf{G}\perp}%
\end{array}
\right)  $ is $4N$.
In addition, $\O$ and ${\bf I}$ are $2N\times 2N$ zero matrix and identity matrix, alternatively.

\begin{figure}[h]
  \centering
  \includegraphics[width=0.4\textwidth]{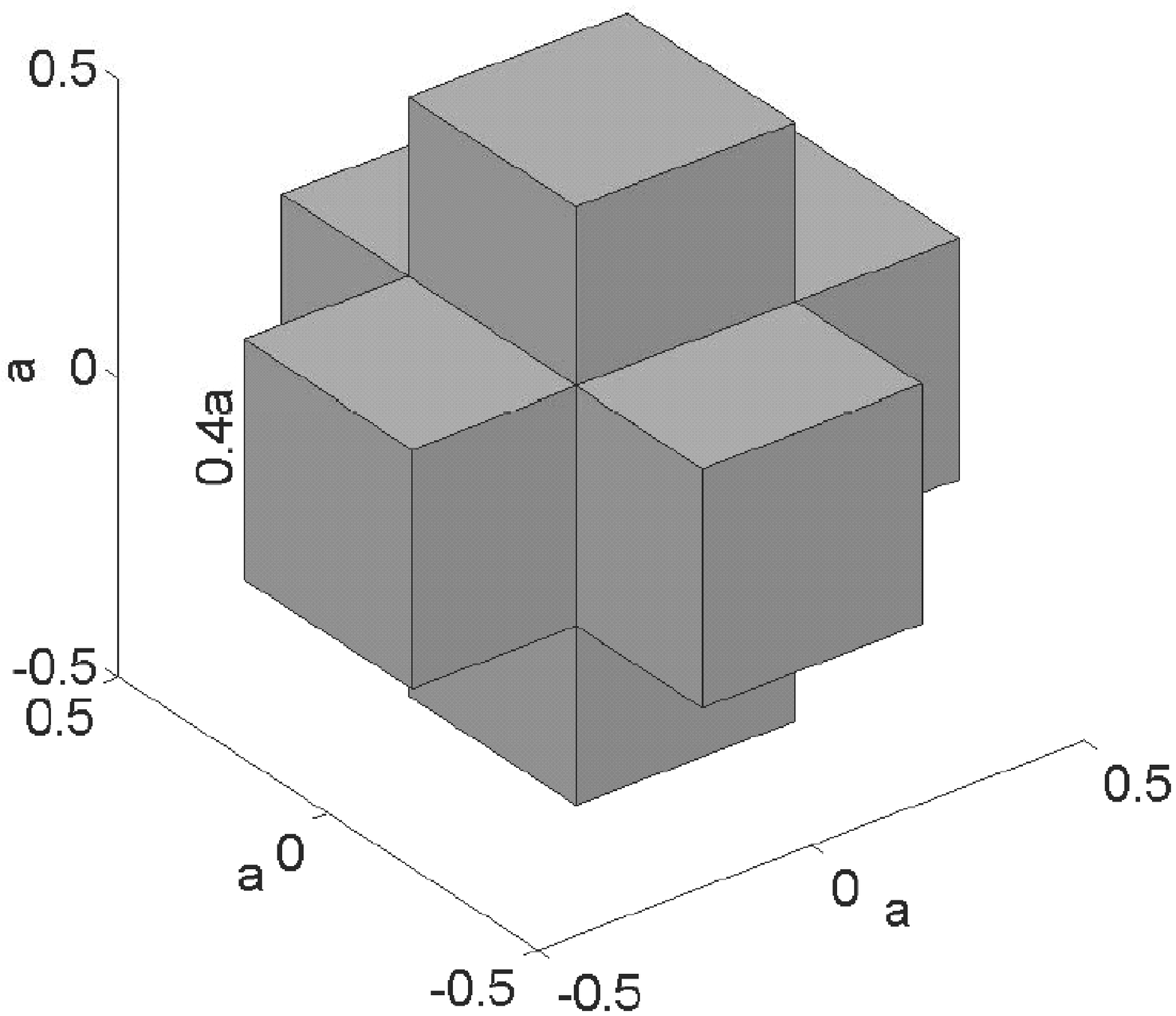}\\
  \caption{The schematic view of a cubic unit cell in which three GaAs square rods cross together from the ${\bf \hat{x}}$, ${\bf \hat{y}}$ and ${\bf \hat{z}}$ direction. The lattice constant, width of square rods and $\epsilon$ of GaAs are $a$, $0.4a$ and $11.43\epsilon_0$, respectively.}\label{fig1}
\end{figure}

For testing this method, we use an Intel centrino 1.4G, 512 MB RAM with matlab
code published on mathworks website to run an isotropic simple cubic case in
which the GaAs square rods --- their widths are $0.4a$, and $a$ is the lattice
constant --- cross together from $\mathbf{\hat{x}}$,$\mathbf{\hat{y}}$ and
$\mathbf{\hat{z}}$ direction in the vacuum. In this system the permittivity
$\epsilon$ of GaAs and vacuum are $11.43\epsilon_{0}$ and $\epsilon_{0}$,
alternatively, and the permeability $\mu$ is $\mu_{0}$ everywhere. You can see
its structure in Fig.(1) and calculation results in Fig.(2). We spent about 6
hours on getting Figs.(2b) and (2c) when using 729 $\{\mathbf{G}\}$ and taking
$17$ $k_{y}$ points from $0$ to $\frac{\pi}{a}$ to accomplish the calculation.
As regards Fig.(2a), it is the band structure which is derived from Eq.(2)
and used to compare with our method. In Fig.(2a), we can find that
$\omega=0.2\frac{2\pi c}{a}$ is not located in band gap, so such kind of
condition should also appeared in our method when we choose the same $\omega$
to plot the contour line or surface. Figure (2b) in which $\omega
=0.2\frac{2\pi c}{a}$ , $k_{z}=0$ and $k_{y}$ scanned from $-\frac{\pi}{a}$ to
$\frac{\pi}{a}$ is the figure of real value solution of $k_{x}$ derived from
Eq.(6). When Fig.(2a) compares with Fig.(2b), we will find out the width of
contour in Fig.(2b) equals the width of $\mathbf{X}\rightarrow{\bf \Gamma}$
 region in Fig.(2a).

\begin{figure}[h]
  \centering
  \includegraphics[width=0.4\textwidth]{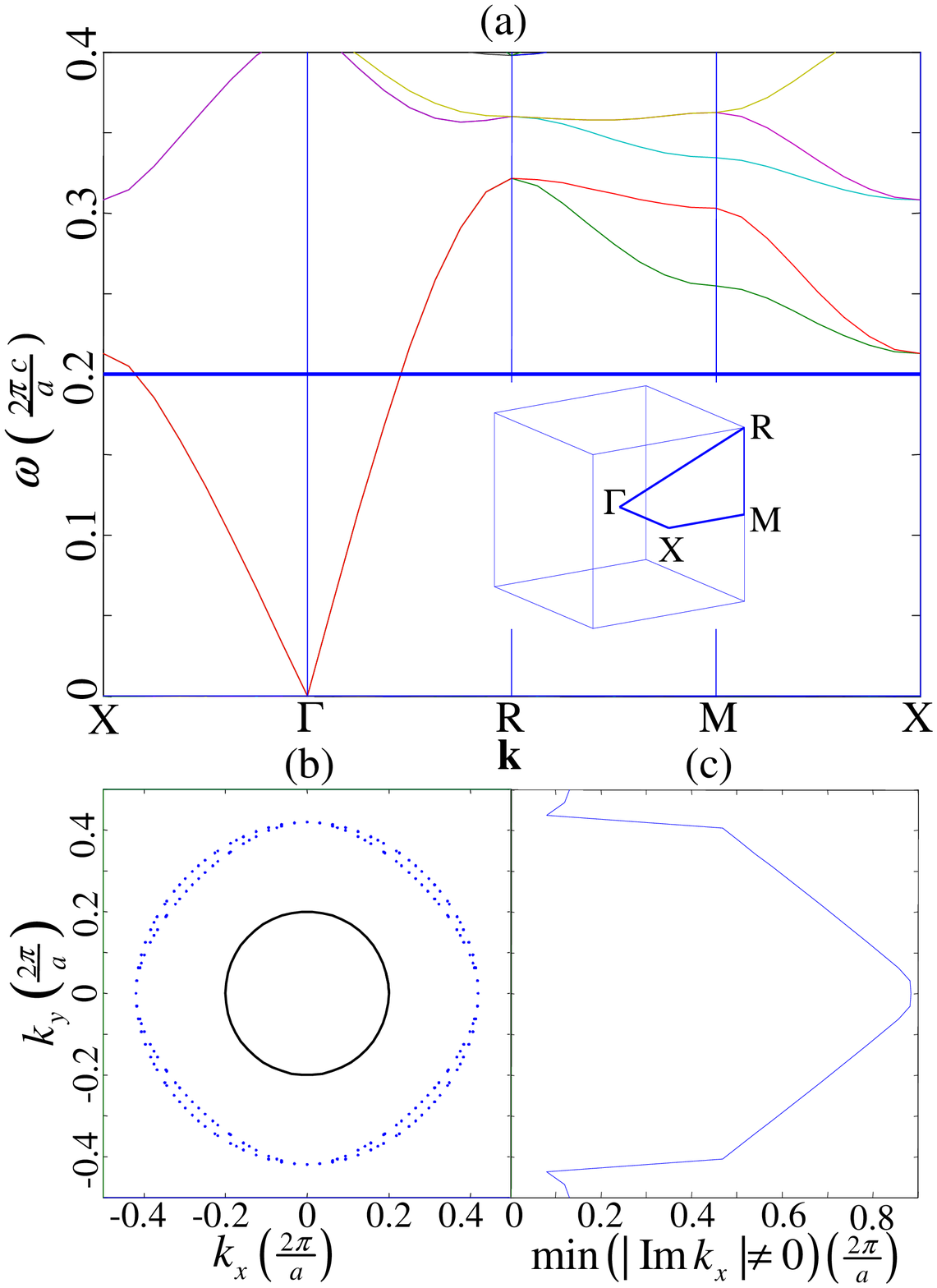}\\
  \caption{The numerical results of Fig.1. (a) is the band structure derived from Eq.(2) and in which the bold line is the $\omega=0.2\frac{2\pi c}{a}$ line. (b) and (c) are the equal frequency contour line of propagation modes in k space and the $\min(|\operatorname{Im}(k_x)|\neq 0)$ vs. $k_y$ figure, alternatively. The circle in (b) denotes the incident light of which $\omega=0.2\frac{2\pi c}{a}$. Both of them are derived from Eq.(6) when $\omega=0.2\frac{2\pi c}{a}$, $k_z=0$ and $k_y$ is scanned from $-\frac{\pi}{a}$ to $\frac{\pi}{a}$.}\label{fig2}
\end{figure}

Besides, we can find that there are two propagation modes toward right when
$k_{y}$ is a fixed number in Fig.(2b). These modes are similar to TE and TM
modes in 2D isotropic PC, however, they can not be distinguished in 3D PC, we
just plot them directly. Furthermore, $C_{4}^{1}$ symmetry exists in Fig.(2b)
but not in the figure of real part of complex $k_{x}$. The reason is the real
number solutions of $k_{x}$ are the $k_{x}$ of the propagation modes which are
the solutions of bulk system in which the $C_{4}^{1}$ symmetry exist. However,
the above is not correct when $k_{x}$ are complex numbers, because the complex
means that there is an interface destroying the $C_{4}^{1}$ symmetry and
facing $\mathbf{\hat{x}}$ direction in the system as well. Therefore, all the
evanescent modes of which $k_{x}$ are complex numbers just exist near the
interface and their penetration depths correspond to $2\pi/\left|
\operatorname{Im}(k_{x})\neq0\right|  $ owing to $e^{i\mathbf{k\cdot r}%
}=e^{i\mathbf{k}^{R}\mathbf{\cdot r}}e^{i\mathbf{k}_{x}^{I}x}$, where $R$ and
$I$ denote real and imaginary parts, alternatively. The most remarkable one of
the complex $k_{x}$ relates to the longest penetration depth denoted as
$\lambda_{LPD}\left(  k_{y},k_{z},\omega\right)  $, because almost nothing but
the propagation modes can exist in this system when the distance from the
detecting position to the interface is larger than $\lambda_{LPD}\left(
k_{y},k_{z},\omega\right)  $. Therefore, a semi-infinite system can be treated
as two individual regions: surface and bulk regions, all the evanescent modes
just exist in the surface region of which the width is $\lambda_{S}$ definded
as $\max\left(  \lambda_{LPD}\left(  k_{y},k_{z},\omega_{0}\right)  \right)
$, where $\omega_{0}$ is a fixed frequency. For a finite size PC, if the
effect of corner is not important, $\lambda_{S}$ decides the smallest size of
PC. If the size is smaller than the smallest one, the system no longer can be
treated as a periodic structured media. Figure(2c) is the figure of
$a/\lambda_{LPD}$ vs. $k_{y}$ at $\omega=0.2\frac{2\pi c}{a}$. This figure
indicates that the $a/\lambda_{LPD}$ drops to zero quickly when $k_{y}$ is
located at the edge of contour in Fig.(2b). This kind of situation arises
while the state located at the edge of contour changes from propagation mode
to evanescent mode.
Besides, because $|k_y|\leq 0.2 \frac{2\pi}{a}$ when the incident light is a propagation mode in vacuum, we can find that $a/\lambda_{LPD}>0.7$. Therefore, the longest penetration depth is $a/0.7$ for all incident light perpendicular to $\mathbf{\hat{z}}$ direction.

In conclusion, because Eq.(6) is a $k_{x}$ eigenvalue equation when $\omega$,
$k_{y}$ and $k_{z}$ are provided, the $\omega$ can be a real number at any
time, and $\epsilon$ and $\mu$ can be the function of $\omega$, $k_{y}$ and
$k_{z}$ or complex tensors. In addition, since most of $k_{x}$ are complex
numbers, the minimum of $\left|  \operatorname{Im}(k_{x})\neq0\right|  $ must
exist, and this value will decide how large a PC is able to treated as a
single crystal if the influence of corner is not important. Therefore, one of
the issues we proceed to research is the influence of corner. 
We thank Prof. Ping Shen for his opinion to excite us to find out the 3D formula EPWE method. 
\section{Appendix}
The $\epsilon_{ij}$ shown as below is the abbreviation of $\epsilon_{{\bf G}-{\bf G}',ij}$. 
\begin{align*}
  {\bf A}=& \left(%
\begin{array}{cc}
  \epsilon^{-1}_{zz} & -\epsilon^{-1}_{zy} \\
  -\epsilon^{-1}_{yz} & \epsilon^{-1}_{yy} \\
\end{array}%
\right),
\end{align*}
\begin{align*}
B_{1,11}=&
  G_x\epsilon^{-1}_{zz}(k+G')_y+(k+G)_z\epsilon^{-1}_{xy}(k+G')_z-
  \\ &G_x\epsilon^{-1}_{zy}(k+G')_z-(k+G)_z\epsilon^{-1}_{xz}(k+G')_y+\omega^2\mu_{yx},\\
B_{1,21}=&
  G_x\epsilon^{-1}_{yy}(k+G')_z+(k+G)_y\epsilon^{-1}_{xz}(k+G')_y-
  \\ &G_x\epsilon^{-1}_{yz}(k+G')_y-(k+G)_y\epsilon^{-1}_{xy}(k+G')_z+\omega^2\mu_{zx},\\
B_{2,11}=&
  G_x\epsilon^{-1}_{zx}(k+G')_z+(k+G)_z\epsilon^{-1}_{xz}G_x'-
  \\ &G_x\epsilon^{-1}_{zz}G_x'-(k+G)_z\epsilon^{-1}_{xx}(k+G')_z+\omega^2\mu_{yy},\\
B_{2,12}=&
  G_x\epsilon^{-1}_{zy}G_x'+(k+G)_z\epsilon^{-1}_{xx}(k+G')_y-
  \\ &G_x\epsilon^{-1}_{zx}(k+G')_y-(k+G)_z\epsilon^{-1}_{xy}G_x'+\omega^2\mu_{yz},\\
B_{2,21}=&
  G_x\epsilon^{-1}_{yz}G_x'+(k+G)_y\epsilon^{-1}_{xx}(k+G')_z-
  \\ &G_x\epsilon^{-1}_{yx}(k+G')_z-(k+G)_y\epsilon^{-1}_{xz}G_x'+\omega^2\mu_{zy},\\
B_{2,22}=&
  G_x\epsilon^{-1}_{yx}(k+G')_y+(k+G)_y\epsilon^{-1}_{xy}G_x'-
  \\ &G_x\epsilon^{-1}_{yy}G_x'-(k+G)_y\epsilon^{-1}_{xx}(k+G')_y+\omega^2\mu_{zz},
\end{align*}
\begin{align*}
C_{1,11}=&
  \epsilon^{-1}_{zz}(k+G')_y - \epsilon^{-1}_{zy}(k+G')_z,\\
C_{1,21}=&
  \epsilon^{-1}_{yy}(k+G')_z - \epsilon^{-1}_{yz}(k+G')_y,\\
C_{2,11}=&
  \epsilon^{-1}_{zx}(k+G')_z + (k+G)_z\epsilon^{-1}_{xz} - \epsilon^{-1}_{zz}(G_x+G_x'),\\
C_{2,12}=&
  \epsilon^{-1}_{zy}(G_x+G_x') - \epsilon^{-1}_{zx}(k+G')_y - (k+G)_z\epsilon^{-1}_{xy},\\
C_{2,21}=&
  \epsilon^{-1}_{yz}(G_x+G_x') - \epsilon^{-1}_{yx}(k+G')_z - (k+G)_y\epsilon^{-1}_{xz},\\
C_{2,22}=&
  \epsilon^{-1}_{yx}(k+G')_y + (k+G)_y\epsilon^{-1}_{xy} - \epsilon^{-1}_{yy}(G_x+G_x'),
\end{align*}
\begin{align*}
D_{11}=&
  (k+G)_z\epsilon^{-1}_{yz} - (k+G)_y\epsilon^{-1}_{zz},\\
D_{12}=&
  (k+G)_y\epsilon^{-1}_{zy} - (k+G)_z\epsilon^{-1}_{yy},
\end{align*}
\begin{align*}
E_{1}=&
  (k+G)_y\epsilon^{-1}_{zy}(k+G')_z + (k+G)_z\epsilon^{-1}_{yz}(k+G')_y-
  \\&(k+G)_y\epsilon^{-1}_{zz}(k+G')_y - (k+G)_z\epsilon^{-1}_{yy}(k+G')_z + \omega^2\mu_{xx},\\
E_{2,11}=&
  (k+G)_y\epsilon^{-1}_{zz}G_x' + (k+G)_z\epsilon^{-1}_{yx}(k+G')_z-
  \\&(k+G)_y\epsilon^{-1}_{zx}(k+G')_z - (k+G)_z\epsilon^{-1}_{yz}G_x' + \omega^2\mu_{xy},\\
E_{2,12}=&
  (k+G)_z\epsilon^{-1}_{yy}G_x' + (k+G)_y\epsilon^{-1}_{zx}(k+G')_y-
  \\&(k+G)_z\epsilon^{-1}_{yx}(k+G')_y - (k+G)_y\epsilon^{-1}_{zy}G_x' + \omega^2\mu_{xz}.
\end{align*}

\end{document}